\documentclass[
    ,final            
  ]
  {aipproc}

\layoutstyle{8x11single}

\usepackage{amsmath}
\usepackage{bm}

\begin{document}

\title{Critical behavior in dislocation systems: power-law relaxation below the yield stress}


\author{P\'eter Dus\'an Isp\'anovity \footnote{E-mail: \texttt{ispanovity@metal.elte.hu}; URL: \texttt{http://dislocation.elte.hu}}}{
  address={Paul Scherrer Institute, CH-5232 Villigen PSI, Switzerland},
  altaddress={Department of Materials Physics, E\"otv\"os University Budapest, H-1518 POB 32, Hungary}
}

\begin{abstract}
Plasticity of two-dimensional discrete dislocation systems is studied. It is shown, that at some threshold stress level the response becomes stress-rate dependent. Below this stress level the stress-plastic strain relation exhibits power-law type behavior. In this regime the plastic strain rate induced by a constant external stress decays to zero as a power-law, which stems from the scaling of the dislocation velocity distribution. The scaling is cut-off at a time only dependent on the system size and the scaling exponent depends on the external stress and on the initial correlations present in the system. These results show, that the dislocation system is in a critical state everywhere we studied below the threshold stress.
\end{abstract}

\maketitle

\section{Introduction}

Plastic deformation of crystalline materials is usually achieved by the motion of lattice dislocations. These linear crystal defects interact via long-range anisotropic stress-fields, and at low temperatures their motion is constrained to a glide plane. In a simplified model used in this paper only parallel edge dislocations are considered with a single glide plane. Although in crystals dislocations usually form complex three-dimensional networks, this model proved capable of reproducing many experimentally observed phenomena related to plasticity, like, e.g.,\ strain avalanche statistics \cite{Miguel2001}, Andrade-creep exponents \cite{Miguel2002, Rosti2010}, and properties of X-ray profiles \cite{Csikor2004, Ispanovity2008b}.

This two-dimensional (2D) model was also found to exhibit a yielding transition, that is, below some yield stress $\tau_\text{y}$ the plastic strain rate decays to zero, otherwise it tends to a constant value \cite{Miguel2002}. This fundamental observation raised the analogy with second-order phase transitions with $\tau_\text{y}$ characterizing the critical point; for a review see \cite{Zaiser2006}. This theory was elaborated by Laurson \emph{et al.}\ by introducing a dynamical correlation length, that diverges as $\tau_\text{y}$ is approached from above \cite{Laurson2010}. In addition, with a cellular automaton technique Zaiser \emph{et al.}\ showed, that as the applied stress approaches $\tau_\text{y}$ from below, both the total accumulated plastic strain and the cut-off of the strain avalanche distribution diverge \cite{Zaiser2005b, Zaiser2006}. According to these results, the concept of a well-defined critical yield point seems to be established for this 2D model.

In this paper it is shown, that a threshold stress $\tau_\text{th}$ can be introduced with the plastic response showing power-law behavior below it, and strong dependence on the applied stress rate above it. It is argued, that below $\tau_\text{th}$ the system behaves like one in a critical state, i.e.,\ the characteristic cut-off times are diverging with increasing system size. This behavior was already confirmed in several different set-ups \cite{Ispanovity2011}. Here it is shown, that if the external stress is applied to the random/uncorrelated system, the exponents of the power-law relaxation change significantly, confirming their dependence on the initial conditions and external stress. According to these results, the concept of the existence of a non-equilibrium critical yield point for this 2D system is challenged, rather criticality for all relaxation states below $\tau_\text{th}$ is suggested.

\section{The dislocation dynamics model}

A set of parallel straight edge dislocations with parallel slip planes is considered with a 2D representation on a plane perpendicular to the dislocation lines. For the motion of dislocations overdamped dynamics is assumed because of the large acting friction forces. By introducing the notations $\bm r_i = (x_i, y_i)$ for the position of the $i$th dislocation, $\bm b_i = s_i(b,0)$ for its Burgers vector ($s_i = \pm 1$ is called its sign), the equation of motion of each dislocation takes the form \cite{Miguel2002}
\begin{equation}
	\dot{x}_i = s_i \left[ \sum_{j=1;\ j\ne i}^{N} \!\!\! s_j \tau_\text{ind}(\bm
r_i - \bm r_j) + \tau_\text{ext}(\bm r_i) \right]\!\!;\qquad \dot{y}_i = 0.
\label{eqn:eq_mot}
\end{equation}
Here $\tau_\text{ind}(\bm r) = \cos (\varphi) \cos (2\varphi)r^{-1}$ is the long-range shear stress field generated by an individual dislocation, $\tau_\text{ext}$ is the external shear stress, and $N$ is the total number of the dislocations in the system. In the rest of this paper the different physical parameters are absorbed in the length-, time-, and stress-scales, as we measure them in the natural units of $\rho^{-0.5}$, $(\rho M Gb^2)^{-1}$, and $Gb\rho^{0.5}$, respectively, where $\rho$ is the dislocation density, $M$ is the dislocation mobility, and $G$ is an elastic constant \cite{Csikor2009}. At the borders of the square-like simulation area periodic boundary conditions are applied.

\section{Yield tests}

\begin{figure}
	\begin{minipage}{\textwidth}
		\begin{center}
		\begin{picture}(0,0)
		\put(41,-22){\sffamily{(a)}}
		\end{picture}
		\includegraphics[angle=-90, scale=0.89]{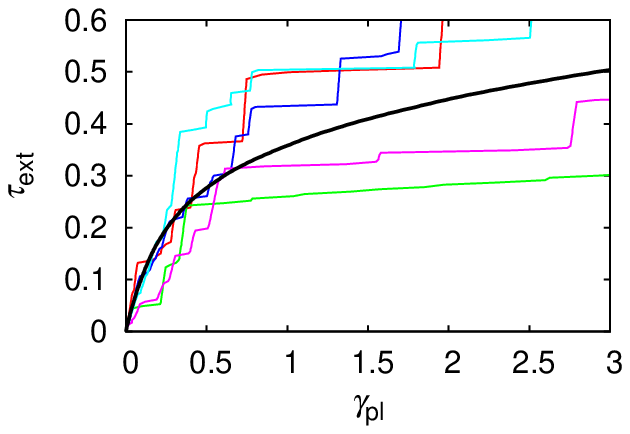} \hspace*{1cm}
		\begin{picture}(0,0)
		\put(41,-22){\sffamily{(b)}}
		\end{picture}
		\includegraphics[angle=-90, scale=0.89]{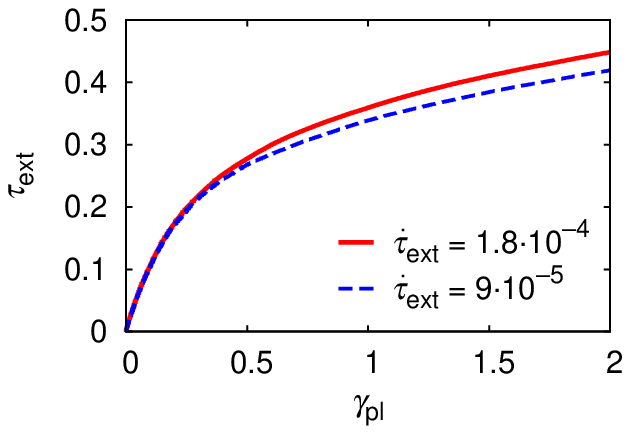}\\
		
		\begin{picture}(0,0)
		\put(41,-22){\sffamily{(c)}}
		\end{picture}
		\includegraphics[angle=-90, scale=0.89]{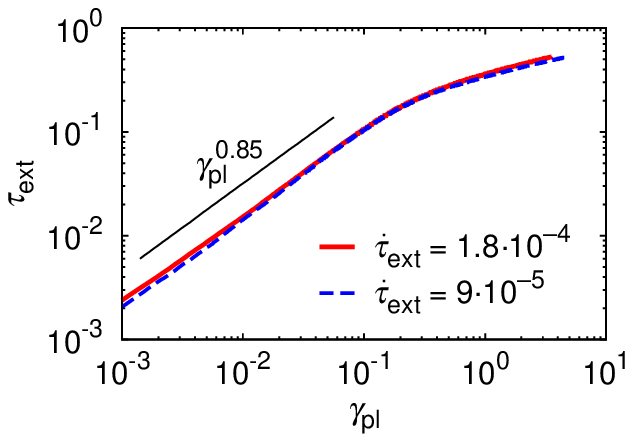} \hspace*{1cm}
		\begin{picture}(0,0)
		\put(41,-22){\sffamily{(d)}}
		\end{picture}
		\includegraphics[angle=-90, scale=0.89]{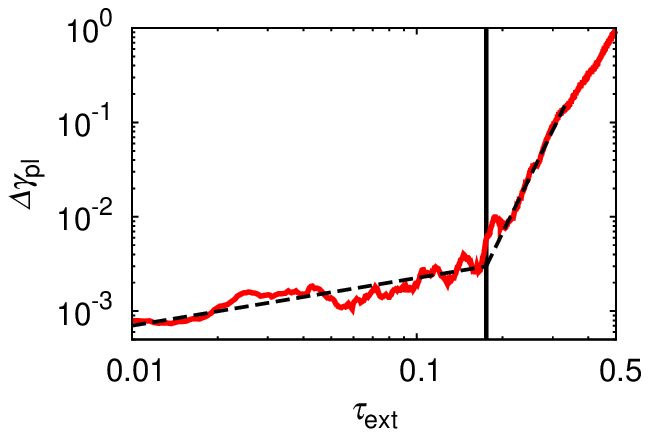}
		\end{center}
	\end{minipage}
	\caption{Yield test with constant stress rate at a system size of $N=128$. (a) Stress-plastic strain curves for different realizations (thick color lines) and the average curve (thin black line). (b) The average stress-plastic strain curves for two different stress rates. (c) The same curves as in (b) now on a double logarithmic plot. The power-law regime for small stresses breaks down around the threshold stress $\tau_\text{th}$. (d) The difference of the average plastic strain values $\Delta \gamma_\text{pl}$ measured at a given stress for the two different stress rates (i.e.,\ it is the difference of the curves in panel (b) after switching the axes).}
	\label{fig:yield}
\end{figure}

In order to study the plastic response of this system stress-controlled yield tests were performed \cite{Ispanovity2010}. The simulations were started from a random arrangement of an equal number of positive and negative sign dislocations (with a total number of $N=128$), which were then let to relax at zero external stress. After this the applied stress $\tau_\text{ext}$ was gradually increased with constant rate. As seen in Fig.~\ref{fig:yield}(a), individual simulations exhibit fluctuating stress-strain curves, with steps corresponding to sudden bursts of activity. One can, however, define an average stress-strain curve over an ensemble by assigning for every applied stress level $\tau_\text{ext}$ the average of the plastic strain $\gamma_\text{pl}$ values measured in the individual simulations [the result is the black thick line in Fig.~\ref{fig:yield}(a)]. It was shown, that for small external stresses this curve is a power-law for at least two orders of magnitude \cite{Ispanovity2010}. At some threshold stress level $\tau_\text{th}$ the power-law relation smoothly breaks down. Further analysis showed, that in the range of this $\tau_\text{th}$ other characteristics, like the plastic strain rate, the fluctuation of the plastic strain, and the coefficient of the inverse cubic tail of the dislocation velocity distribution also behave similarly \cite{Ispanovity2010}. According to numerical fitting, an approximate value of $\tau_\text{th} \approx 0.17$ was suggested \cite{Ispanovity2010}. It should be emphasized, however, that since the observed power-law breakdown is smooth $\tau_\text{th}$ is not characterizing a single point, rather a transition regime.

In Fig.~\ref{fig:yield} results about the stress rate dependence of the plastic response are reported. The averaging procedure for the stress-plastic strain curve shown in Fig.~\ref{fig:yield}(a) and described above was repeated for simulations with a halved stress rate of $\dot{\tau}_\text{ext} = 9 \cdot 10^{-5}$ (in natural units). The result in Fig.~\ref{fig:yield}(b) shows that the two average stress-strain curves overlap for small stresses, and split for larger stresses. As expected, at a given $\tau_\text{ext}$ larger strains are observed for the smaller rate. The double logarithmic plot of Fig.~\ref{fig:yield}(c) [with the same curves as in Fig.~\ref{fig:yield}(b)] confirms the power-law type behavior below the threshold stress $\tau_\text{th}$. Figure~\ref{fig:yield}(d) plots the difference of the two curves of Fig.~\ref{fig:yield}(b) with switched axes, i.e., $\Delta \gamma_\text{pl}(\tau_\text{ext}$) denotes the difference of the two average plastic strain values $\gamma_\text{pl}$ measured for the two strain rates at a given $\tau_\text{ext}$. (Note that in \cite{Ispanovity2010} the notation of $\Delta \gamma$ was used for a different quantity, the strain fluctuations.) It is seen, that $\Delta \gamma_\text{pl}$ also exhibits a transition at the threshold stress $\tau_\text{th}$, and the system is much more sensitive to the driving rate above $\tau_\text{th}$ than below. In conclusion, the threshold stress introduced separates two distinctly different regimes, presumably marking a yielding phenomenon.

\section{Relaxation tests}

In this section the power-law regime below the threshold stress $\tau_\text{th}$ is investigated. The following different simulation scenarios were suggested for this purpose: (i) relaxation of initially randomly distributed dislocations to an equilibrium state at zero external stress, (ii) response of the equilibrated system to an inserted fixed dislocation, and (iii) response of the equilibrated system to an external stress below $\tau_\text{th}$ \cite{Ispanovity2011}. In all three cases relaxation to a steady state is observed with the following properties \cite{Ispanovity2011}:
\begin{itemize}
	\item The mean absolute velocity of the dislocations $|v(t)|$ in all scenarios and the plastic strain rate $\dot{\gamma}_\text{pl}(t)$ in scenario (iii) decay to zero as power-law. This scaling regime is cut-off at a time $t_1$ only dependent on the system size as $t_1 \sim \sqrt{N}$, and not depending on the external stress.
	\item In the scaling regime both the symmetric $P_\text{s}$ and the antisymmetric $P_\text{a}$ part of the dislocation velocity distribution exhibit scaling property as follows:
	\begin{equation}
		P_\text{s}(v,t) = t^\alpha f(t^\alpha v) \quad \text{ and } \quad P_\text{a}(v, t) = t^\gamma g(t^\beta v)
	\label{eqn:scaling}
	\end{equation}
	with appropriate $f$ and $g$ scaling functions. The power-law dependence of the mean absolute velocity and the plastic strain rate can be directly deduced from the above scaling of $P_\text{s}$ and $P_\text{a}$, respectively: $|v(t)| \sim t^{-\alpha}$ and $\dot{\gamma}_\text{pl}(t) \sim t^{\gamma-2\beta}$ \cite{Ispanovity2011}. In the last expression $2\beta - \gamma$ is called Andrade exponent.
	\item The exponents are different in each case suggesting that they depend on the initial conditions and that, therefore, they are not determined only by the type of the interactions and dynamics of the system.
\end{itemize}

\begin{figure}
	\begin{minipage}{\textwidth}
		\begin{center}
		\begin{picture}(0,0)
		\put(41,-28){\sffamily{(a)}}
		\end{picture}
		\includegraphics[angle=-90, scale=0.89]{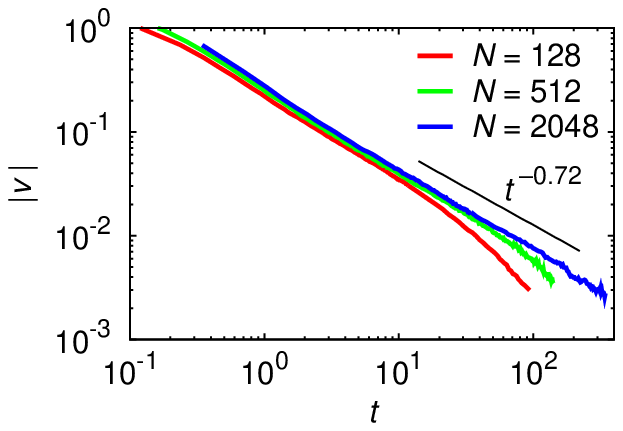} \hspace*{1cm}
		\begin{picture}(0,0)
		\put(142,-22){\sffamily{(b)}}
		\end{picture}
		\includegraphics[angle=-90, scale=0.89]{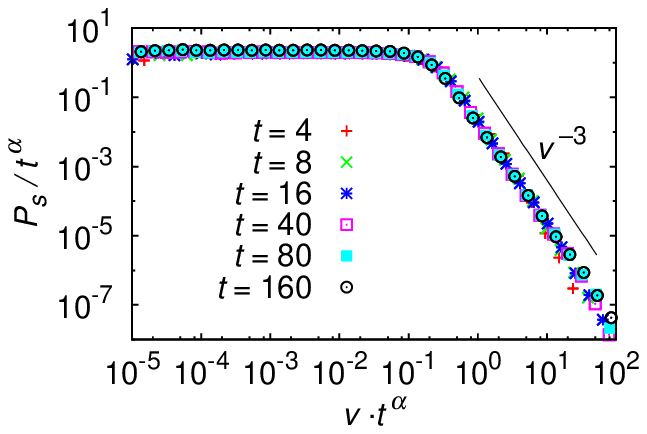}\\
		
		\begin{picture}(0,0)
		\put(41,-22){\sffamily{(c)}}
		\end{picture}
		\includegraphics[angle=-90, scale=0.89]{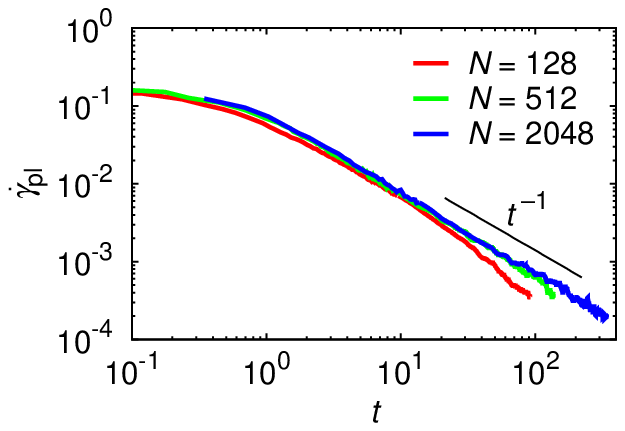} \hspace*{1cm}
		\begin{picture}(0,0)
		\put(142,-22){\sffamily{(d)}}
		\end{picture}
		\includegraphics[angle=-90, scale=0.89]{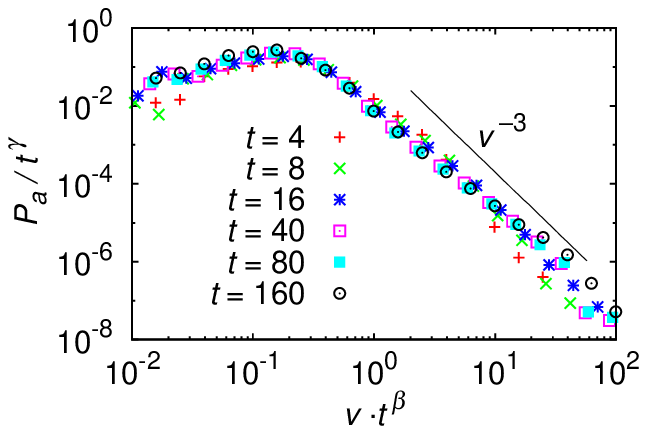}
		\end{center}
	\end{minipage}
	\caption{Relaxation from a random initial state with an applied stress of $\tau_\text{ext} = 0.17$. (a) Evolution of the mean absolute velocity for different system sizes $N$. (b) Scaling of the symmetric part of the velocity distribution $P_\text{s}$ as Eq. (\ref{eqn:scaling}) with $\alpha = 0.72$ for the system size of $N=2048$. (c) Evolution of the plastic strain rate $\dot{\gamma}_\text{pl}$ for different system sizes $N$. (d) Scaling of the antisymmetric part of the velocity distribution $P_\text{a}$ as Eq. (\ref{eqn:scaling}) with $\beta = 0.75$ and $\gamma = 0.5$ for the system size of $N=2048$.}
	\label{fig:relax}
\end{figure}

Here the last point is elaborated, by repeating simulation scenario (iii) (response to a small external stress) with the stress applied not on an initially relaxed configuration, but on a completely random system. The applied external stress is $\tau_\text{ext}=0.17 \approx \tau_\text{th}$ (at smaller stresses similar behavior is found). It is seen in Fig.~\ref{fig:relax}(a) that the mean absolute velocity decays as $|v(t)| \sim t^{-\alpha}$ with an exponent of $\alpha \approx 0.72$. The cut-off time behavior is identical to the one found before and described above. As in the other cases, the scaling of $P_\text{s}$ is observed [Eq.~(\ref{eqn:scaling})] in the scaling regime with the same $\alpha$ exponent [Fig.~\ref{fig:relax}(b)]. Power-law decay characterizes $\dot{\gamma}_\text{pl}(t)$, too [Fig.~\ref{fig:relax}(c)], also accompanied by the scaling of $P_\text{a}$ as in Eq.~(\ref{eqn:scaling}) with $\beta \approx 0.75$ and $\gamma \approx 0.5$ [Fig.~\ref{fig:relax}(d)]. The Andrade exponent of the strain rate is therefore $2\beta -\gamma \approx 1.0$. Note, that in \cite{Csikor2009} with the same simulation set-up a stretched exponential form was obtained for $\dot{\gamma}_\text{pl}(t)$, presumably because of too small system sizes.

It is interesting to compare the observed exponents with the previous results. For $\alpha$ here $0.72$ is found, while in scenario (i) (i.e.,\ the same simulation with zero external stress) $0.85$ was observed. So the scaling exponent is changed by only modifying the level of the external stress. On the other hand, when comparing the simulation of Fig.~\ref{fig:relax} with scenario (iii) (stress applied on a relaxed system) the only difference is in the initial configuration (relaxed with internal correlations or random). The exponents are again clearly different: $\beta$ changes from $0.5$ to $0.75$ and $\gamma$ from $0$ to $0.5$. The corresponding Andrade-exponent of $\dot{\gamma}_\text{pl}(t)$ is also shifted from around $0.6$ to around $1.0$. The exponents, therefore, depend both on the level of the external stress and the statistical properties of the initial configuration.

\section{Discussion}

It was shown, that below the threshold stress this 2D dislocation system exhibits slow relaxation with a cut-off time diverging with the system size. This behavior is typical for systems being in a critical point, it is, therefore, concluded that this system is always critical in this regime. Consequently, description of the yielding transition with a single critical point is challenged. In addition, the scaling exponents are sensitive to the initial conditions and to the external stress level. In particular, the Andrade-exponent $2\beta - \gamma$ of the plastic strain-rate $\dot{\gamma}_\text{pl}(t)$ is not universally $2/3$ in this system, but can be as large as $1$ depending on the properties of the starting configuration and the external stress.

Further questions such as how this critical behavior is changed above the threshold stress and how $\tau_\text{th}$ relates in fact to the yield stress $\tau_\text{y}$ discussed in the Introduction still remain to be addressed.

\begin{theacknowledgments}

The author would like to thank G.\ Gy\"{o}rgyi and I.\ Groma for fruitful conversations and for commenting on the manuscript. The financial support of the Hungarian Scientific Research Fund (OTKA) under Contract No.\ K 67778 and The European Union and the European Social Fund under the Grant Agreement No.~T{\' A}MOP-4.2.1/B-09/1/KMR-2010-0003 are also gratefully acknowledged.

\end{theacknowledgments}

\bibliographystyle{aipproc}
\bibliography{paper_ispanovity}

\end{document}